\documentclass[MSNbibl,nameyear,seceqn,dvips]{arxstspdf}
\usepackage{flushend}
\usepackage{stfloats}
\usepackage{graphicx}

\volume{29}
\issue{1}
\pubyear{2014}
\firstpage{91}
\lastpage{94}
\doi{10.1214/13-STS460} 

\makeatletter
\newcommand{\EE}{\mathbb{E}}
\renewcommand{\citep}[1]{(\citeauthor{#1}, \citeyear{#1})}
\renewcommand{\citet}[1]{\citeauthor{#1} (\citeyear{#1})}
\newproclaim{Locationmodel}{Location model with conjugate Gaussian prior}
\makeatother

\begin{document}
\begin{frontmatter}

\title{Discussion of Big Bayes Stories and~BayesBag}
\runtitle{Discussion and BayesBag}

\begin{aug}
\author[a]{\fnms{Peter} \snm{B\"uhlmann}\corref{}\ead[label=e1]{buhlmann@stat.math.ethz.ch}}
\runauthor{P. B\"uhlmann}

\affiliation{ETH Z\"urich}

\address[a]{Peter B\"uhlmann is Professor, Seminar for Statistics, ETH Z\"
urich, CH-8092 Z\"urich,
Switzerland \printead{e1}.}

\end{aug}


%

\end{frontmatter}

\section{Introductory Remarks}

I congratulate all the authors for their insightful papers with wide-ranging
contributions. The articles demonstrate the power and elegance of the
Bayesian inference
paradigm. In particular, it allows to incorporate prior knowledge as well
as hierarchical model building in a convincing way. Regarding the
latter, the contribution by Raftery, Alkema and German is a very
fascinating piece, as it addresses a set of problems of great
public interest and presents
predictions for the
world populations and other interesting quantities with
uncertainty regions. Their approach is based on a hierarchical model,
taking various characteristics into account (e.g., fertility
projections). It
would have been very difficult to come up with a ``better'' solution which
would be as clear in terms of interpretation (in contrast to a ``black-box
machine'') and which would provide (model-based) uncertainties for the
predictions into the future.

\section{Uncertainty, Stability and Bagging the~Posterior}

Many of the papers quantify in one or another form various notions of
uncertainties. In the Bayesian framework, this is usually based on the
posterior distribution. An old ``debate'' is how much the results are
sensitive to the choice of the prior, and I believe that some reasonable
sensitivity analysis can lead to much insight. The sensitivity with respect
to ``perturbed data'' though is not easily captured by the Bayesian
framework. In the context of prediction, Leo Breiman (\citeauthor{brei96},
\citeyear{brei96}, \citeyear{brei96b})
has pointed to issues of
stability with respect to perturbations of the data, \citet{bous02}
provide some
mathematical connections to prediction performance while
Meinshausen and B{\"u}hlmann
(\citeyear{mebu10})
present some theory and methodology for controlling the frequentist error
of expected false positives.

As an example, the (frequentist) Lasso \citep{tibs96} is very unstable for
estimating the
unknown \mbox{parameters} in a linear model, in particular, if the correlation
among the covariates is high (for two highly correlated variables where at
least one of them has a substantially large regression coefficient, the
Lasso selects either one or the other in an unstable fashion). Thus, the
MAP for a Gaussian linear model with a Double-Exponential prior for the
regression coefficients is \mbox{unstable}. The posterior distribution is probably
more stable but, presumably, it is still ``rather'' sensitive with respect
to perturbation of the data: if the data would look a bit different, the
posterior might be ``rather'' different. The situation becomes more exposed
to stability problems when using spike and slab priors (Mitchell and Beauchamp, \citeyear{mitchbeau88}),
due to increased sparsity.\looseness=-1

We can stabilize the posterior distribution by using a bootstrap and
aggregation scheme, in the spirit of bagging \citep{brei96b}. In a
nutshell, denote by $\mathcal{D}^*$ a bootstrap- or subsample of the data
$\mathcal{D}$. The posterior of the random parameters $\theta$ given
the data
$\mathcal{D}$ has c.d.f. $F(\cdot|\mathcal{D})$, and we can
stabilize this using
\[
F_{\mathrm{BayesBag}}(\cdot|\mathcal{D}) = \EE^* \bigl[F \bigl(\cdot|\mathcal{D}^*
\bigr) \bigr],
\]
where $\EE^*$ is with respect to the bootstrap- or subsampling
scheme. We call it the \emph{BayesBag} estimator. It can be approximated
by averaging over $B$ posterior
computations for bootstrap- or subsamples, which might be a rather
demanding task (although say $B=10$ would already stabilize to a certain
extent). Note that when conditioning on the data, the posterior
$F(\cdot|\mathcal{D})$ is a fixed c.d.f., but when taking the view
point that
the data could change, it is useful to consider randomized perturbed
versions $F(\cdot|\mathcal{D}^*)$ which are to be aggregated.

%
\begin{table*}
\tablewidth=9.7cm
\tabcolsep=0pt
\caption{2.5\% and 97.5\% quantiles of the posterior $F(\cdot
|\overline{X}_n)$
in (\protect\ref{posterior})
and\break of the BayesBag (bagged posterior) in (\protect\ref{bag-Gauss}). The data was\break
generated once
using a single realized value of $\theta= 1.31$}\label{tab1}
\begin{tabular*}{9.7cm}{@{\extracolsep{\fill}}lcc@{}}
\hline
\textbf{Sample size} & \multicolumn{1}{c}{$\boldsymbol{(2.5\%,97.5\%)}$ \textbf{posterior}}
& \multicolumn{1}{c@{}}{$\boldsymbol{(2.5\%,97.5\%)}\ \boldsymbol{\mathit{BayesBag}}$}
\\
\hline
$n=1$ & $(-0.69, 2.81)$ & $(-1.30, 3.41)$
\\
$n=10$ & $(0.10, 1.32)$ & $(-0.16, 1.56)$
\\
\hline
\end{tabular*}
\end{table*}

The following simple and rather stable example shows that such a
bagging scheme outputs a larger uncertainty which is perhaps more
appropriate.

\begin{figure}[b]

\includegraphics{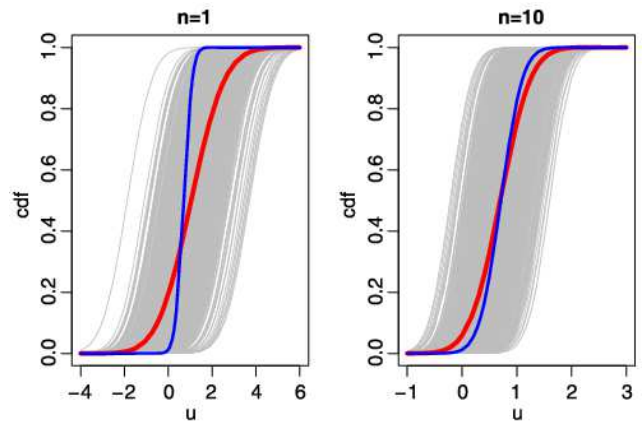}

\caption{1000 bootstrapped cumulative distribution functions
$F(u|\overline{X}_n^*)$ of $\theta|
\overline{X}_n^*$. The BayesBag (i.e., mean) $\EE^*[F(u|\overline
{X}_n^*)]$ in
(\protect\ref{bag-Gauss}) (thick red line) and the cumulative distribution function
$F(u|\overline{X}_n)$ of the classical posterior of $\theta|\overline{X}_n$
in (\protect\ref{posterior}) (blue line). Left
panel for $n=1$ and right panel for $n=10$, and note the different scales
for the x-axis. The data is as in Table~\protect\ref{tab1}.}\label{fig1}
\end{figure}

\begin{Locationmodel*} Consider the model
\[
\begin{array}{r@{\quad}l}
 &\theta\sim\mathcal{N} \bigl(0,\tau^2 \bigr),
\\[2pt]
\mbox{conditional on } \theta{:}&  X_1,\ldots,X_n
 \,\mathrm{i.i.d.} \sim \mathcal{N} \bigl(\theta,\sigma^2 \bigr).
\end{array}
\]
It is well known that the posterior distribution equals
\[
\theta|\overline{X}_n \sim\mathcal{N} \biggl(\frac{\sum_{i=1}^n X_i}{n +
\sigma^2/\tau^2},
\biggl(\frac{1}{\tau^2} + \frac{n}{\sigma^2} \biggr)^{-1} \biggr).
\]
Denote by $F(\cdot;\overline{X}_n)$ the c.d.f. of the
posterior distribution, that is,
%
\begin{eqnarray}
\label{posterior}
&&F(u;\overline{X}_n) = \Phi \biggl(u,\mathrm{mean}=
\frac{n \overline
{X}_n}{n +
\sigma^2/\tau^2},
\nonumber
\\[-8pt]
\\[-8pt]
&&\phantom{\hspace*{80pt}}\mathrm{var} = \biggl(\frac{1}{\tau^2} + \frac{n}{\sigma^2}
\biggr)^{-1} \biggr),\nonumber
\end{eqnarray}
where $\Phi(u,\mathrm{mean} = m,\mathrm{var} = s^2) = \Phi
((u-m)/s)$ and
$\Phi(\cdot)$ denotes the c.d.f. of $\mathcal{N}(0,1)$. We can
either use the
nonparametric bootstrap, with resampling the data with replacement, or a
parametric
bootstrap (assuming here that $\sigma^2$ is known):
%
\begin{equation}
\label{boot-param} X_1^*,\ldots, X_n^* \,\mathrm{i.i.d.}\,
\mathcal{N} \bigl(\hat{\theta},\sigma ^2 \bigr),\quad \hat{\theta} =
\overline{X}_n.
\end{equation}
With the parametric bootstrap in (\ref{boot-param}), we can easily
calculate the \emph{BayesBag} estimator:
%
\begin{eqnarray}
\label{bag-Gauss} &&\EE^* \bigl[F \bigl(u;\overline{X}_n^* \bigr) \bigr]
\nonumber
\\
&&\quad= \int \Phi \biggl(\frac{u-r}{\sqrt{({1}/{\tau^2} + {n}/{\sigma^2})^{-1}}} \biggr)
\nonumber
\\[-8pt]
\\[-8pt]
&&\phantom{\quad= \int}{}\cdot\varphi \biggl(r,
\mathrm{mean} = \frac{n \overline{X}_n}{n + \sigma
^2/\tau^2},\nonumber
\\
&&\phantom{\hspace*{62pt}}\mathrm{var} = \frac{n\sigma^2}{(n + \sigma^2/\tau^2)^2} \biggr) \,dr,\nonumber\nonumber
\end{eqnarray}
where $\varphi(r,\mathrm{mean} = m,\mathrm{var} = s^2) = s^{-1}
\varphi((r
- m)/s)$ and $\varphi(\cdot)$ denotes the p.d.f. of $\mathcal
{N}(0,1)$. We
consider the posterior credible region by computing the 2.5\% and 97.5\%
quantiles of $F(\cdot;\overline{X}_n)$ and we compare these quantiles with
the corresponding ones from the \mbox{BayesBag} $\EE^*[F(\cdot;\overline{X}_n^*)]$
above in (\ref{bag-Gauss}). We only
consider here the case with $\sigma^2=1$ and $\tau^2 = 4$, and the
results are
given in Table~\ref{tab1}.
Of course, we can also look at the variability of the posterior via the
bootstrapped c.d.f.'s $F(\cdot|\overline{X}_n^*)$, instead of considering
the bootstrap mean (\emph{BayesBag}) only. Figure~\ref{fig1} illustrates
that variability can
be rather high, but the situation obviously improves as sample size
increases.
\end{Locationmodel*}

It is worth pointing out that, in general, one could use a parametric
bootstrap when using $\hat{\theta}$ as the MAP of the posterior
distribution, and such a scheme could be used in models with
complex hierarchical and dependence structures.

The frequentist approach usually does not address stability issues either
and, in addition, assigning $p$-values and
confidence intervals in complex scenarios is a nontrivial
problem. Recent progress has
been achieved for high-dimensional sparse models
(\cite{minnieretal11}; \cite{zhangzhang11}; \cite{bogetal13}; \cite{pb13};
\cite{vdgetal13}, cf.); regarding
the issue of constructing ``stable $p$-values,'' an approach based on
subsampling and appropriate aggregation of $p$-values is described in
\citet{memepb09}. Yet, much more work in frequentist inference would be
needed to cope with, for example, high-dimensional hierarchical models in
non-i.i.d. settings such as space--time processes or clustered data, or, as
another example, the population dynamic model in the beautiful
paper by Kuikka, Vanhatalo, Pulkkinen, M\"antyniemi and Corander in
this issue.

\section{Networks and Graphical Models}

The paper by Johnson, Abal, Ahern an Hamilton presents an interesting
application by using Bayesian inference for a Bayesian network (as is well
known, the term ``Bayesian network'' does not require Bayesian
inference at
all---and it is somewhat confusing). The arrows in the directed acyclic
graph often indicate causal relations (\cite{pearl00}; \cite{sgs00}) and, as such,
the model allows for causal conclusions. Great care is needed, of course,
when the DAG is misspecified or estimated from observational data since
causal conclusions are depending in a very ``sensitive way'' on the
underlying DAG. A lot of work exists regarding identifiability of the DAG
from observational data (\cite{pearl00}; \cite{sgs00};
\cite{shpipea08}; \cite{hoy09}; \cite{petbu13}, cf.),
and, obviously, there
are ill-posed situations such as with a bivariate Gaussian distribution
where one cannot
identify the causal direction between two variables.
In the Bayesian framework, the problem of identifiability does not
exist in
a ``direct sense'': but I believe it must come in through another channel,
presumably by a high sensitivity with respect to prior specifications.
Due to severe identifiability problems, causal inference based on
observational data is ill-posed or depends on nontestable
assumptions. However, one can nevertheless (under some assumptions) derive
lower bounds on absolute values of causal effects \citep{makapb09}. As
lower bounds, they are conservative and a Bayesian average bound would be
interesting.

In view of nontestable assumptions, causal models should be validated
with randomized experiments. Often though, this cannot be done due to
limited \mbox{resources} or ethical reasons. The field of molecular biology with
simple organisms is an interesting application where causal
model validation is feasible thanks to gene knock-out or other manipulation
methods. We pursued this in the past, for estimated causal structures and
models based on frequentist approaches, for the organisms yeast
\citep{mapb10} and
arabidopsis thaliana \citep{steketal12}. These two papers
indicate that it is indeed possible to predict to a certain extent
lower bounds
of causal strength and relations
based on observational (and very high-dimensional) data. Such a
conclusion can only be made post-hoc, after validation---and
validation has
nothing to do whether a Bayesian or any other inference machine has been
used.

\section*{Acknowledgment}

I thank Nicolai Meinshausen for interesting
comments and suggesting the name \emph{BayesBag}.


%

\end{document}